\documentstyle[12pt,color,epsfig,subfigure]{article}
\advance\textheight by \topskip
\newcommand{\bea}{\begin{eqnarray}}
\newcommand{\eea}{\end{eqnarray}}

\def\gtap{\raisebox{-.4ex}{\rlap{$\sim$}} \raisebox{.4ex}{$>$}}
\def \etslash{\not \! E_T }

\begin{document}

\begin{center}

{\Large \bf 
Enriching the exploration of the mUED 
model with event shape variables at the CERN LHC} \\
\vspace*{1cm}
\renewcommand{\thefootnote}{\fnsymbol{footnote}}
{ {\sf Amitava Datta$^1$}, {\sf Anindya Datta${}^{2}$},
 and {\sf Sujoy Poddar${}^{3}$}
} \\
\vspace{10pt}
{\small ${}^{1}$} 
{\em  Indian Institute of Science Education and Research, Kolkata, \\
          Mohanpur Campus, PO: BCKV Campus Main Office,\\
	            Mohanpur - 741252, India} \\
   ${}^{2)}$ {\em Department of Physics, University of Calcutta,
92 A.P.C. Road, Kolkata 700 009, India}  \\
   ${}^{3)}$ {\em Department of Physics, 
            Netaji Nagar Day College,\\
             170/436, N.S.C. Bose Road, Kolkata - 700092, India
          }\\
\normalsize

\abstract 

\noindent
{We propose a new search strategy based on the event shape variables
for new physics models where the separations among the masses of the
particles in the spectrum are small. Collider signature of these
models, characterised by low $p_T$ leptons/jets and low missing $p_T$,
are known to be difficult to look for. The conventional search strategies
involving hard cuts may not work in such situations. As a case study,
we have investigated the hitherto neglected jets $+$ missing $E_T$
signature -known to be a challenging one - arising from the pair
productions and decay of $n =1$ KK-excitations of gluons and quarks in
the minimal Universal Extra Dimension (mUED) model. Judicious use of
the event shape variables, enables us to reduce the Standard Model
backgrounds to a negligible level . We have shown that in mUED,
$R^{-1}$ upto $850 ~\rm GeV$, can be explored or ruled out with 12
fb$^{-1}$ of integrated luminosity at the 7 TeV run of the LHC. We
also discuss the prospects of employing these variables for searching
other beyond standard model physics with compressed or partially
compressed spectra.}\\

PACS numbers: { \tt 12.60.-i, 04.50.cd, 13.85.-t} 
\end{center}

\section{Introduction}

One of the main goals of the ongoing LHC experiment at CERN is to find
out any new dynamics that could be operative at the energy scale of
Tera electron volts (TeV) among the elementary particles. Apart from the
search of Higgs boson, both the ATLAS and the CMS experiments are
engaged in looking for the signals of scenarios beyond the
Standard Model. Among these, models defined in one or more space-like
extra dimensions need special attention.  These models can be divided
broadly into two classes. In models proposed in \cite{ADD} and
\cite{RS}, all the Standard Model~(SM) fields are confined in a $1 +3$
dimensional sub-space of a larger space-time manifold, while the
Gravitational interaction can perceive the full space time
manifold. After compactification of the extra space like dimensions,
the effective four dimensional theory consists of towers of gravitons
interacting with SM fields.  However, we are interested in a class of
models wherein some or all of the SM fields can access the extended
space-time manifold \cite{acd, cms}.  Such
extra-dimensional scenarios could lead to a new mechanism of
supersymmetry breaking \cite{antoniadis1}, relax the upper limit of
the lightest supersymmetric neutral Higgs
\cite{relax}, address the issue of fermion mass hierarchy 
\cite{Arkani-Hamed:1999dc}, provide a cosmologically
viable dark matter candidate \cite{Servant:2002aq}, interpret the
Higgs as a quark composite leading to a successful EWSB without the
necessity of a fundamental scalar or Yukawa interactions
\cite{Arkani-Hamed:2000hv}, and lower the unification scale down to a
few Tev \cite{dienes, blitz}. Our concern here is a particularly
interesting framework, called the minimal Universal Extra Dimension
(mUED) scenario, characterized by a single flat extra dimension,
compactified on an $S^1 /Z_2$ orbifold (with radius of
compactification, $R$)\cite{acd}. This extra space like dimension is
accessed by all the SM particles. From a 4-dimensional viewpoint,
every field in the SM will then have an infinite tower of Kaluza-Klein
(KK) modes, each mode being identified by an integer, $n$, called the
KK-number. The zero modes ($n=0$) are identified as the
corresponding SM states. The orbifolding is essential to ensure that
fermion zero modes have a chiral representation. But it has other
consequences too. {\em First}, the physical region along the extra
direction $y$ is now smaller [$0, \pi R$] than the periodicity [$0,
2\pi R$], so the KK number ($n$) is no longer conserved. What remains
actually conserved is the even-ness and odd-ness of the KK states,
ensured through the conservation of KK parity, defined by
$(-1)^n$. {\em Secondly}, Lorentz invariance is also lost due to
compactification, and as a result the KK masses receive bulk and
orbifold-induced radiative corrections \cite{cms, rad_corr}. The bulk
corrections are finite and nonzero only for bosons. The orbifold
corrections, which vary logarithmically with the cutoff ($\Lambda$),
depend on group theoretic invariants, as well as Yukawa and quartic
scalar couplings of the gauge and matter KK fields and hence are
flavor-dependent. This induces a mass splitting among the different
flavors of the same KK level, further to what has already been caused
by the different zero mode masses. The model thus can be described by
{\em two dimensionful parameters}, namely the inverse of
compactification radius, $R^{-1}$ and the cut-off scale, $\Lambda$. We
will not present the expressions for the radiatively corrected masses
of the different KK-modes of the SM particles.  However, these can be
easily obtained from \cite{BDMR}. Independent of the values of the
input parameters, the lightest among the $n=1$ KK states turns out to
be $\gamma^1$, the $n =1$ KK-excitation of photon.  Typically, if
$R^{-1} = 500$ GeV, mass of $\gamma^1$ is slightly above 500 GeV, just
above lie the KK leptons ($L^1$, $\nu ^1$) and weak bosons ($W^{\pm
1}$, $Z^1$) in the region of 500-550 GeV, further up are the KK quarks
($Q_{L,R}^1$) near 600 GeV, and at the peak the KK gluon, $G^1$, (the
heaviest) hovers around 650 GeV.

Conservation of the KK-parity ensures the lightest KK particle (LKP)
is stable (hence being a natural candidate for the Dark-matter
~\cite{Servant:2002aq}) and that the level-one KK-modes would
be produced only in pairs.  This also ensures that the KK modes do not
affect electroweak processes at the tree level. And while they do
contribute to higher order electroweak processes, in a loop they
appear only in pairs resulting in a substantial suppression of such
contributions, thereby allowing for relatively smaller KK-spacings.
In spite of the infinite multiplicity of the KK states, the KK parity
ensures that all electroweak observables are finite (up to
one-loop)\cite{db}\footnote{The observables start showing cutoff
sensitivity of various degrees as one goes beyond one-loop or
considers more than one extra dimension.}, and a comparison of the
observable predictions with experimental data yields bounds on the
compactification radius $R$.  Constraints on the UED scenario from the
measurement of the anomalous magnetic moment of the muon \cite{nath},
flavour changing neutral currents \cite{chk}, $Z \to b\bar{b}$ decay
\cite{santa}, the $\rho$ parameter \cite{acd,appel-yee}, several other
electroweak precision tests \cite{ewued}, yield $R^{-1}~\gtap~300$
GeV.

The fact that such a small value for $R^{-1}$ (equivalently, small KK
spacings) is still allowed, renders collider search prospects very
interesting both in the context of hadronic \cite{collued, kong1, BDMR, kirtiman_ad_jhep, kirtiman_biplob_ued}
 and leptonic
\cite{ILC, kong} colliders.

At the very outset it was realized that the signatures  of the mUED   
model at hadron colliders has an inherent problem \cite{cms}. The signature 
with the largest cross-section at hadron colliders is the jets + 
missing transverse momenta ($\etslash$) which is similar to the 
traditional squark- gluino signal in supersymmetric (SUSY) models. There 
is, however, an important difference.

It has been already mentioned above that the spectrum of mUED is very much
compressed. As a result, the transverse
momenta/energy spectra of all the visible particles - the missing
transverse momenta spectrum included- are soft. Consequently the
conventional search strategies to dig out the signals of mUED from the
SM backgrounds using strong cuts on visible/missing $p_T$ are not very
efficient. Such cuts on the other hand are the most potent tools in
the arsenal of the SUSY hunter.

Subsequently the viability of jets $+ \etslash$ channel has never been explored in the framework of 
mUED, because of the general belief that the signal of mUED in this
channel will be overwhelmed by the QCD background.  All the earlier
analyses in the context of mUED, in fact, are either based on search
of $n = 2$ KK-excitations \cite{kong1, kong} of SM particles or on the $n=1$
KK-excitations giving rise to multi-leptons in association with jets
and $\etslash$ \cite{BDMR, kirtiman_ad_jhep, kirtiman_biplob_ued}. The
bulk of the collider events stemming from such model remain
unexplored till date.

In this work we focus on this hitherto neglected channel. Moreover our 
analysis will be restricted to the search prospects at the ongoing
experiments at 7 TeV. It would be 
important to mention here, that both the ATLAS and the CMS 
collaborations have looked for the above  jets $+ 
\etslash$ signature \cite {SUSY-CMS, SUSY-ATLAS} using the accumulated 
data of 1.04 fb$^{-1}$ from the current LHC run at 7 TeV.  In
principle, these analyses could be used to constrain the mUED
parameters. However, the CMS/ATLAS analyses are aimed for SUSY models
motivated by the minimal gravity mediated SUSY breaking (mSUGRA),
where, the masses of the sparticles are well separated over most of
the parameter space. As a result high $p_T$ jets/leptons and a hard
$\etslash$ spectrum is expected in the signal. Thus the search
strategies of the LHC collaborations involve hard cuts on $p_T$ and
$\etslash$ to suppress the huge SM backgrounds (including QCD). For
example, only those events are retained which have $\etslash$
greater than 100 GeV.  Moreover, the leading jet is required to have
$p_T$ greater than 100 GeV.

We shall show the distributions of $\etslash$ and the $p_T$ of the 
leading jet for a representative mUED model in a later section. They  
will indicate unambiguously that the signatures of this model cannot
survive the hard cuts usually employed by the LHC collaborations. Thus 
it is quite possible that the signatures of the mUED model remain 
buried in current LHC data.   

It should emphasized that this is a generic problem (not specific to
mUED only) which confronts the search strategy for any model having a
compressed mass spectrum. For example, in an unconstrained minimal
supersymmetric standard model (MSSM) it is quite possible that the
entire sparticle spectrum is quite compressed. Based on various
theoretical motivations, models with partially compressed mass spectra
have also been proposed \cite{comprsd-SUSY, stealth-SUSY}.  It would
be interesting to device an alternative search strategy for such
scenarios.

In this article we will show that judicious use of the event shape
variables (defined below) would be very efficient in reducing huge SM
background from QCD, $t \bar t$ and $W/Z +$ jets events confronting
the jets + $\etslash$ signal.  Using this new strategy, we could also
push up the sensitivity of the current LHC experiments to the
parameters of the mUED model compared to an earlier analysis using the
kinematic variable $M_{T2}$ \cite{nojiri}.

Before delving into the analysis let us briefly discuss the processes
and the relevant decay cascades that contribute to the signal.  We
will confine to the production of $n = 1$ KK-level excitations only.
These particles can only be produced in pairs by the virtue of
KK-parity conservation.  In LHC, the colliding partons being the
gluons or quarks, pair production of $Q^{1}_{L,R} Q^{1}_{L,R}$, $G^{1}
G^{1}$, $G^{1} Q^{1}_{L,R}$ would be highly enhanced and these
processes contribute to our signal significantly. Once produced,
$G^{1}$ will decay to a $Q^{1}_{L,R} $ along with a SM quark
($Q^{0}_{L,R} $) with equal probabilities. $Q^{1}_{R}$ only can decay
to $Q^{0}_{R}$ and the LKP ($\gamma^1$) .  On the other hand,
$Q^{1}_{L,} $ decays to $W^{\pm 1}$ or $Z^{1}$ (with Brs. $2\over3$
and $1\over3$ respectively) with a SM quark.

It may be recalled that $Z^{1}$ or $W^{\pm 1}$ does not decay
hadronically.  $Z^{1}$ decay results either into $\nu {\bar \nu}
\gamma^1$ (with Br. of 0.5) or into $l_L {\bar l_L} \gamma^1$ (with
Br. of 0.16 for each lepton flavour). On the other hand, $W^{1}$
decays into $l \nu \gamma^1$ (with Brs of 0.33 for each lepton
flavour). It must be emphasised here, that decay patterns and 
branching fractions of $n = 1$ KK-mode fields are independent of the 
mUED model parameters.

Following the above discussions one can see that the $G^{1} G^{1}$ production is the source of 4 jets, while
$G^{1} Q^{1}$ ($Q^{1} Q^{1}$) production leads to 3 (2) jets at the
parton level.  In addition, $\tau$ (coming from $W^1$/ $Z^1$) decay in
hadronic channels will also contribute to our signal enhancing the
number of jets at the parton level itself.  Consequently, the pair
production of $n=1$ KK-gluons and quarks, would most of the time end
up in producing jets $+ \etslash$ final state.  Demanding leptons in
the final state would necessarily mean that production of $Q_L^{(1)}$s
are only being considered and we are throwing away the dominant part
of the cross-section involving productions of $Q_R^{(1)}$s.
 
 All the previous analysis of mUED signal at the LHC were done with
 multi-lepton final state, which necessarily have a smaller
 (effective) signal cross-section.  Of course, there is one advantage
 using the leptonic final states. The SM background rate for the
 multi-lepton final state is also moderate and easy to tame with more
 conventional kinematic cuts used in new particle searches. However,
 as already mentioned all kind of signals arising from a particular
 new physics model must be looked for. Throwing away a class of
 signatures which has the largest cross-section, makes the search
 incomplete.
 
 In this work we have taken a strategy which remove this
 incompleteness and utilizes the large cross-section of jets $+
 \etslash$ final state. The SM background in this channel (arising
 from QCD production of jets, $t \bar t$ production, $W/Z + jets$
 production) is undoubtedly challenging and orders of magnitude larger
 than the signal.  Kinematic cuts, like lower cuts on the $p_T$ of
 particles in the final state or $\etslash$, which are generally used
 for new particle searches, are of not very effective in reducing the
 backgrounds. At this juncture the event shape variables, namely,
 $\alpha_T$ and $R_T$, play a crucial role in taming these huge
 backgrounds without affecting the signal too much.
 
 In the next section we will in detail describe our analysis with
 emphasis om the event-shape variables.  However, before delving into
 the detail, a few features on the parameters of the mUED model need
 our attention. Existing collider and other low energy experimental
 data allow values of $R^{-1}$ to be higher than 300 GeV.  On the
 other hand, the analysis of relic density of LKP dark matter sets an
 upper limit of 700 GeV according to \cite{LKP_dark_matter}.  However,
 we will not be restricted by this upper limit in the following
 analysis and will try to see how much one can push up the search
 limit with the 7 Tev run of LHC.

\section {Analysis and Results}
At the LHC, total production cross-section of $G^{1} G^{1}$,
$G^{1} Q^{1}$, $Q^{1} Q^{1}$ pairs are 0.03 pb, 0.66 pb and
1.21 pb respectively at the leading order (LO ) for $R^{-1} = 700$ GeV with $\Lambda R
= 40$. In the absence of any next to leading order (NLO) QCD corrections to the pair production cross-sections of 
strongly interacting  $n=1$ KK- excitations in mUED, we have
used only the LO signal cross-sections in our analysis. It is also
worth noting that the NLO corrections to the lowest order QCD dijet
cross-section is also not known. If the K-factor arising from the NLO
corrections to the signal cross-section is approximately the same as
that for the overall background, $S /\sqrt{ B}$ will increase by
$\sqrt{K}$. Since K is expected to be $\geq$ 1, the NLO cross-section
is likely to give a better significance. On the other hand using a
typical value of K = 1.5 for the signal, we find that even if the over all K-factor
of the background is 3, the significance computed from the LO cross
section will reduce by 0.9. Thus the estimates based on the LO cross
sections are likely to be fairly conservative.

Signal cross-sections are estimated with the Pythia -6.4.20 \cite{PYTHIA} using the LO CTEQ6L  parton distribution functions (PDF) \cite{CTEQ6}, setting both the scales of PDF and 
$\alpha_s$ at $ \sqrt{\hat s}$ where $\hat {s} $ being the partonic CM energy. 
 The dominant SM backgrounds those can
give rise to jets $+ \etslash$ energy signature are $t \bar t + $
jets, $W/Z +$ jets, QCD production of jets. The sub-dominant
contributions come from $WW +$ jets, $WZ +$ jets and $ZZ +$ jets
productions.  $t \bar t$ production and QCD production of jets have
been estimated using Pythia, while cross-sections for the W/Z
productions have been calculated using ALPGEN \cite{ALPGEN} in
conjunction with Pythia \footnote{The cross-sections for W/Z $+$
n-jets, WW/ZZ/WZ $+$ n-jets ($ n = 1,2$) have been calculated using
ALPGEN subjected to the initial selection cuts of $p_T>20$ GeV,
$|\eta|\le 4.5$ and the jet-jet separation, $\Delta R (j,j)
>0.5$. These cross-sections then were fed into Pythia for parton
showering and to include the ISR/FSR effects.}.  The cross-section for
QCD events have been computed by Pythia in two bins: (a) $25 ~\rm GeV
< \sqrt{\hat {s}} < 400 ~\rm GeV$ (denoted by QCD1 in Table.1) and (b) $400
\rm~ GeV < \sqrt{\hat {s}} < 1000 \rm ~GeV $ (denoted by QCD2 in Table.1)
.
The contributions from other bins being negligible will not be shown any further.
In our simulation using Pythia we have taken into account the effects
of initial and final state radiation as well as fragmentation and hadronization.
A simple toy calorimeter simulation has been implemented with the
following criteria:

\begin{itemize}

\item The calorimeter coverage is $\vert \eta \vert < 4.5$ with
segmentation
of $\Delta \eta \times \Delta \phi = 0.09 \times 0.09$ which
resembles a generic LHC detector.

\item A cone algorithm with $\Delta R$ = $\sqrt {\Delta\eta^2 +
\Delta\phi^2}= 0.5 $ has been used for jet finding.

\item Jets are ordered in
$E_T$ with $E^{jet}_{T,min} = 20$ GeV.
\end{itemize}
Here, $\eta$ and $\phi$ are the pseudo-rapidity and azimuthal angle of the respective visible 
objects.

\begin{figure}[tb]
\begin{center}
\includegraphics[width=\textwidth]{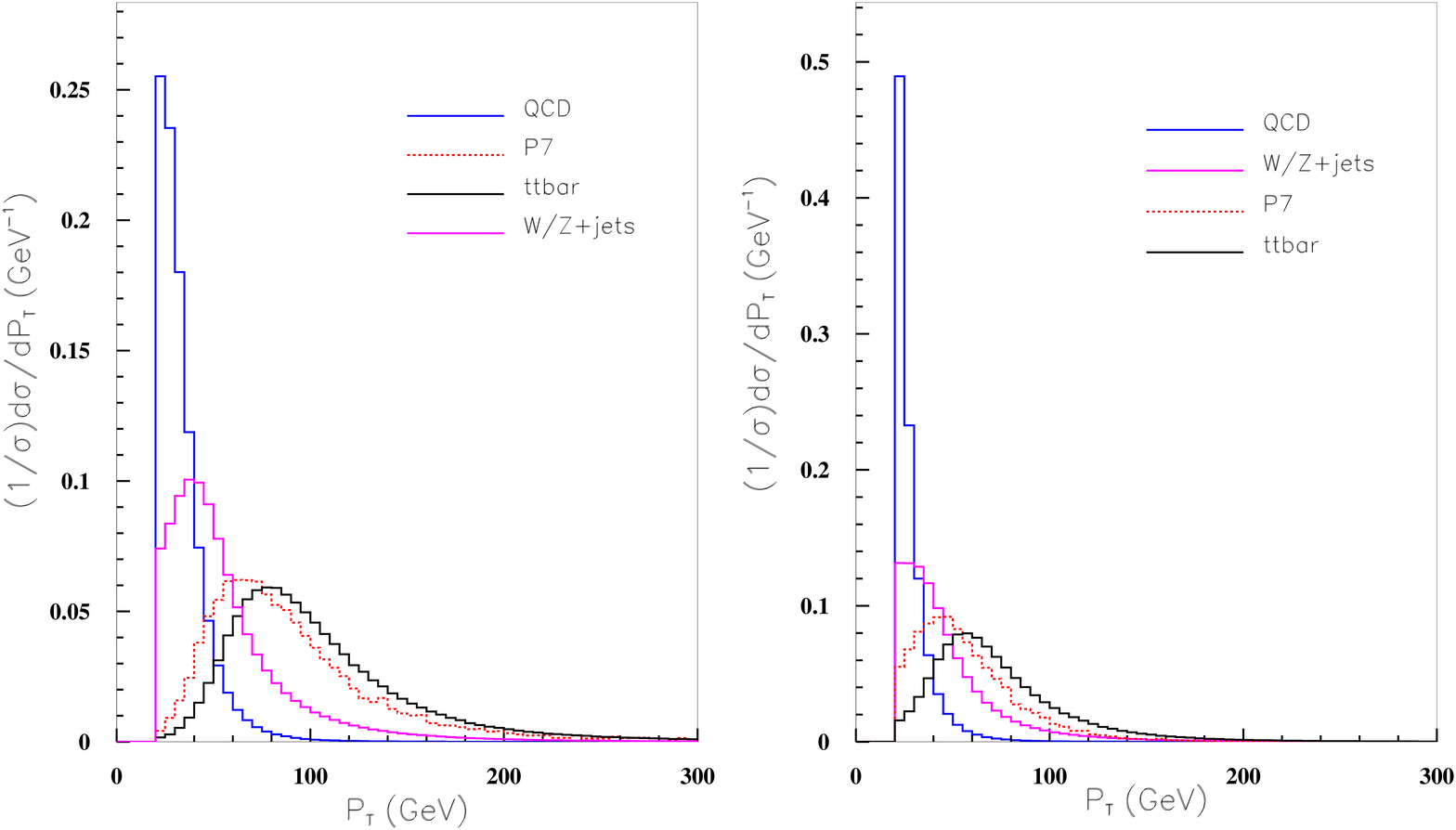}
\includegraphics[width=\textwidth]{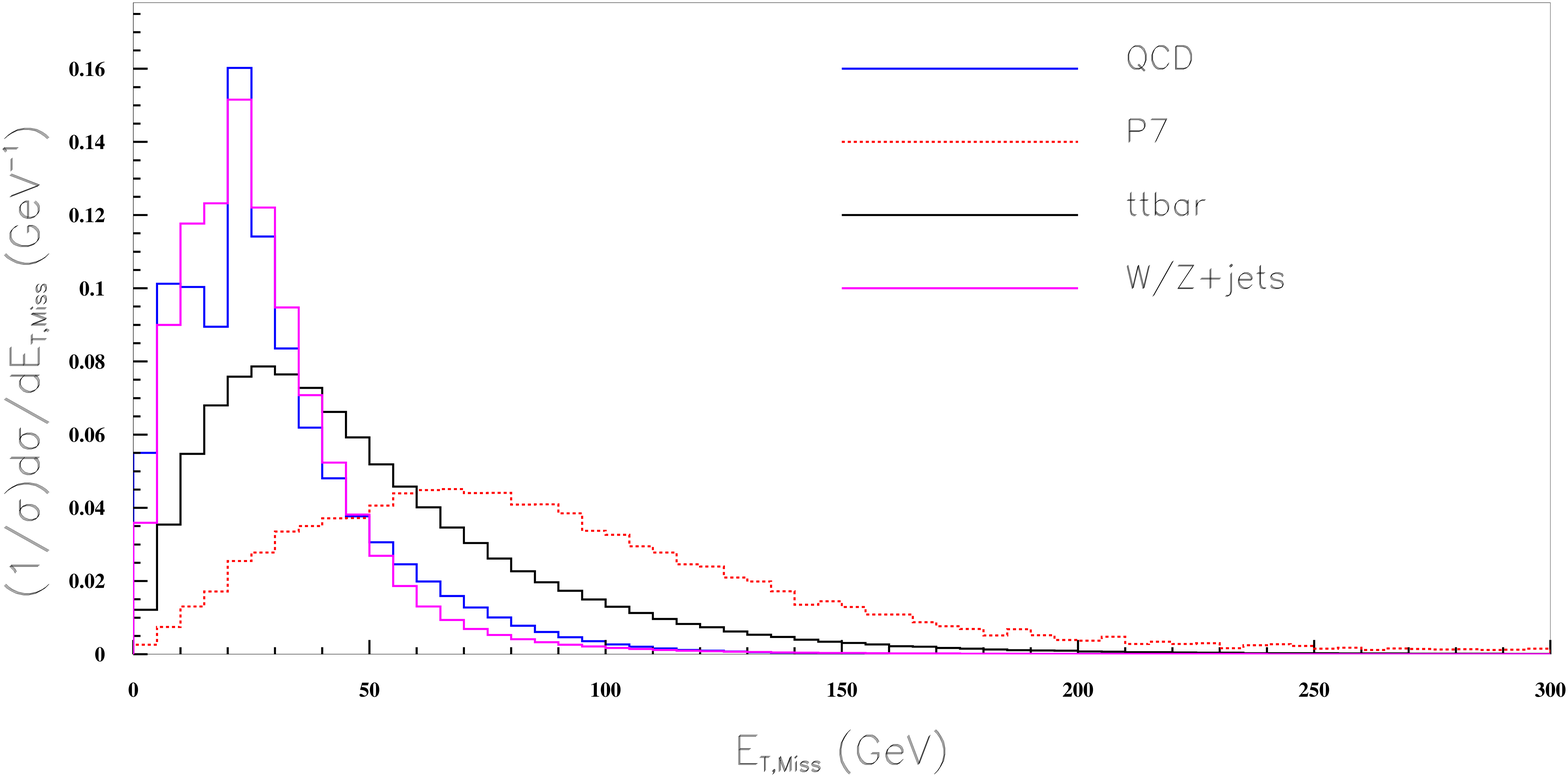}
\end{center}
\caption{
Normalised $p_T$ distributions of two highest $p_T$ jets (upper panles) and normalised missing
$E_T$ distributions of signal and SM backgrounds (lower panel) . In the fiigures, P7 denotes signal with $R^{-1} = 700$ GeV and $\Lambda R = 10$.
}
\label{dist_etmis}
\end{figure}

The total background cross-section overwhelms the signal by several 
orders of magnitude. So one needs to choose some judicious set of cuts 
to enhance the signal to background ratio.  Dominant, SM backgrounds do 
not have real source of missing energy (i.e. neutrinos).  Apparent $p_T$ 
imbalance arises from the finite detector resolution and mis-measurement 
of jet energies in the detector.  Thus one may think that using a rather 
hard cut on $\etslash$ could tame the SM backgrounds for the jets $+ 
\etslash$ signature. However, due an highly compressed mUED mass 
spectrum, jets (in general any visible SM particle) coming from the 
decay of KK-quarks and gluons in case of the signal are quite soft, 
producing a rather soft visible $p_T$ spectra, which in turn gives rise 
to a soft $\etslash$ spectrum.  To demonstrate this, we have 
plotted the $p_T$ distributions of two leading jets and the $\etslash$ 
in Figs.\ref{dist_etmis} for signal (with $R^{-1} = 700$ GeV and $\Lambda R = 
10$ ) and dominant SM backgrounds. One can see from the Figs. that for 
both the signal and SM processes, above distributions peak around rather 
low values of the respective kinematic variables.  Consequently, one 
cannot require events with high $p_T$ (typically $p_T ^j > 100$) 
\cite{SUSY-CMS, SUSY-ATLAS}. Rejection of hard leptons in the final 
state would further restrict our control in reducing the SM background.


In such a situation (events with low missing energy and no lepton), {\em 
event-shape variables}, namely $R_T$ \cite{Guchait} and $\alpha_T$ 
\cite{randall-alpha-T}, are known to be very useful. The CMS 
collaboration has used the variable $\alpha_T$ for controlling the 
background while looking for the signature of SUSY from the jets $+ 
\etslash$ data at the 7 TeV run of LHC. It has also been shown recently 
in \cite{Guchait}, that the SM backgrounds to SUSY signals can be 
brought down to a negligible size by using $R_T$ at the LHC.

The {\em event-shape} variable, $R_T$, is defined by: $$ R_T = \frac
{\Sigma _1 ^{n_j^{min}} \; p_T ^{j_i}}{H_T} $$ where $H_T$ is defined
to be the scalar sum of $p_T$ of all jets.  Here, $n_j^{min}$ denotes
the required minimum number of jets satisfying the criteria : $p_T > 40 ~\rm
GeV $ and $ \vert \eta_j \vert \leq 3$.
 
In fact, $R_T$ gives us a control over the number and hardness of the
reconstructed jets simultaneously. In our case, signal events are
mainly comprised of 2/3/4 partonic jets, which justifies our choice of
($n_j^{min} =$) 3 leading jets in defining (the numerator of ) $R_T$.

The variable $\alpha_T$ is defined as the ratio of the $p_T$ of the
second hardest jet to the invariant mass of the two highest $p_T$ jets
\cite{randall-alpha-T} and is well known to be very potent in reducing
the QCD di-jet events in particular.
 
\begin{figure}[tb]
\begin{center}
\includegraphics[width=\textwidth]{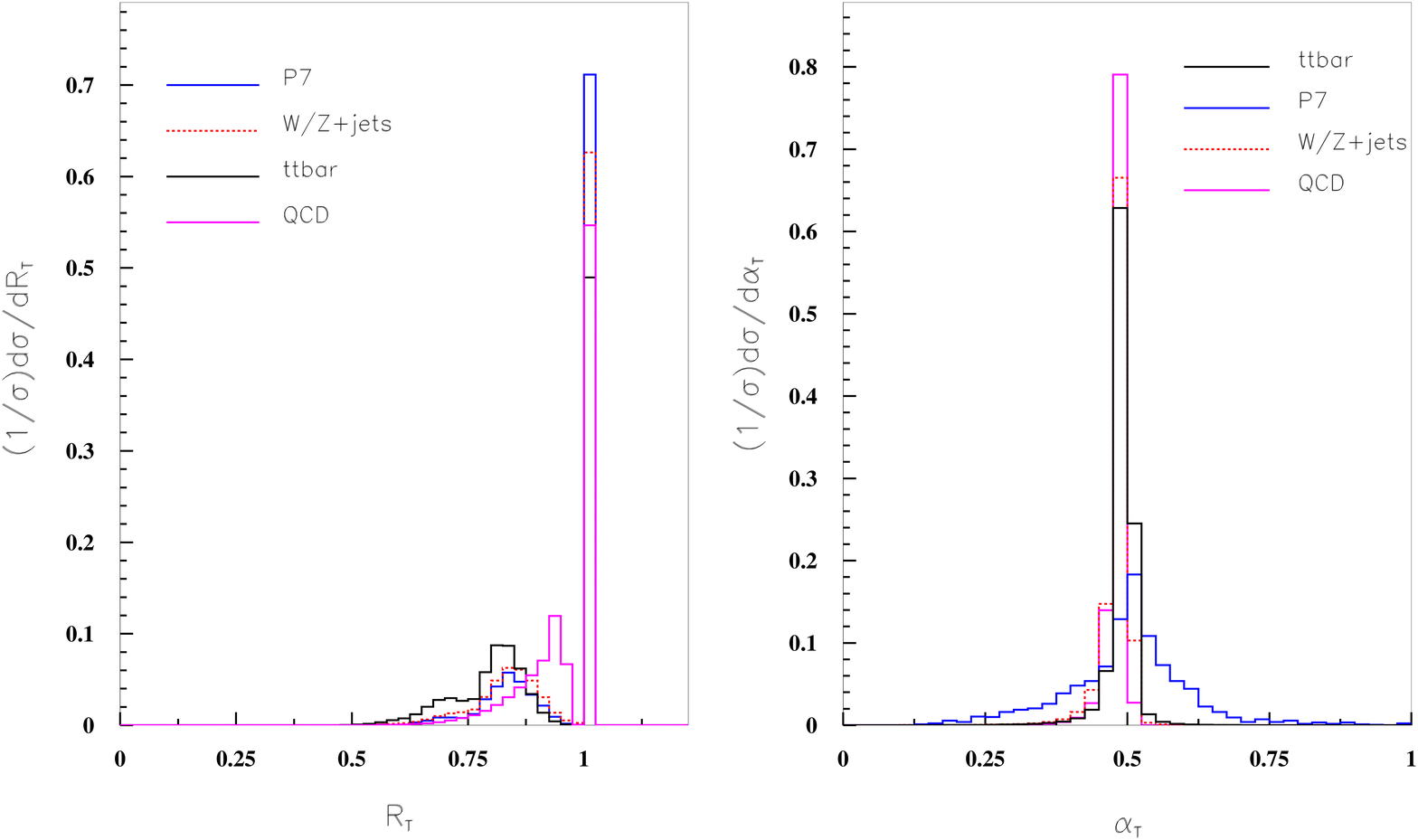}
\end{center}
\caption{
Normalised $R_T$ (left-panel) and $\alpha_T$ (right-panel)
distributions of signal and SM backgrounds.  In the fiigures, P7 denotes signal with $R^{-1} = 700$ GeV and $\Lambda R = 10$.
}
\label{dist_event-shape}
\end{figure}

To demonstrate the usefulness of $R_T$ and $\alpha_T$, we will plot
the distributions of these variables for signal and backgrounds in
Fig. \ref{dist_event-shape}.  It is evident from $R_T$ and $\alpha_T$
distributions in Fig.\ref{dist_event-shape}, that a judicious choice
of these variables can isolate the signal events from the backgrounds.
  

We have implemented following cuts in succession to enhance the signal to background ratio.

\begin{itemize}
\item {\bf C1:} No isolated lepton ($e$, $\mu$) with $p_T > 10$ GeV and $|\eta| <2.5$ are required. Isolated leptons are identified with the criterion 
$\Delta R (l,j) >0.5$, where $\Delta R (l,j)$ denotes the separation between
a lepton($l$) and a jet ($j$) in the $\eta-\phi$ plane.

\item {\bf C2:} Events with $\etslash >50$ GeV are selected.

\item {\bf C3:}   Events with  $R_T \le 0.8$ only are selected.

\item {\bf C4:} Events with $H_T>400$ GeV and $\alpha_T >0.60$ 
(discussed earlier) are selected. 

\item {\bf C5:}  $b$-jet identification has been performed in our analysis 
according to the following procedure. A reconstructed jet with $|\eta|< 2.5$ 
corresponding to the coverage of tracking detectors matching with a $B$-hadron 
of decay length $> 0.9$ mm has been marked $tagged$. 
This criteria ensures that single $b$-jet tagging efficiency
(i.e., the ratio of tagged $b$-jets and the number of taggable $b$-jets)
$\epsilon_b \approx 0.5$ in $t \bar t$ events. Finally in our signal we have 
required {\em the signal to be free from tagged $b$-jet events}.     

\end{itemize}

We note in passing that a trigger $H_T >$ 250 GeV like the one
employed by the CMS collaboration in their $\alpha_T$ analysis
\cite{cms_trigger} of jets + $\etslash$ signal can be quite efficient
for our signal. However, it should be stressed that for a model where
the particle spectrum is not compressed $\alpha_T$ is one of the many
variables which can distinguish the signal and the background. In fact
both CMS and ATLAS collaborations have analysed LHC data without using
the eventshape varibles and, in the context of mSUGRA for example,
have obtained stronger constraints.  In contrast for models with
compressed spectra the options are rather limited and $\alpha_T$
and/or other event shape variables may be invaluable for establishing
the signal.

\begin{table}[!htb]
\begin{center}\

\begin{tabular}{|c|c|c|c|c|c|c|c|}
       \hline
        &  $\sigma$(pb)& $N_{EV}$ & C1 & C2 & C3 & C4 & C5\\
       \hline
        P7 & 1.9  & 0.1M &59778 &53814 &2169 & 153 & 130 \\
       \hline
       QCD1 & $8.6 \times 10^7$ &50M &49885275&336450 &207 & 0 & 0 \\
       \hline
       QCD2 & 1775.0&  8M  &7984488 &2161548 &88093 &0 &0\\
       \hline
       $t \bar t$ & 56.8 & $ 1M$ &621233 &183320 &29288 &17  &*\\
       \hline
       $W +1j$ & 13390& 5M &  4088569 &217476 &1241 &0 &0\\ 
       \hline
       $W +2j$ & 3073  &3M & 2448188 & 252165 & 5726 & 0 & 0 \\ 
       \hline
       $Z +1j$ & 4235 &4M &3674020 &275566 &1036& 0&0 \\ 
       \hline
       $Z +2j$& 970 &1M &918306 & 128750 & 2387 &0 &0 \\ 
       \hline

      \end{tabular}
       \end{center}
       
\caption{Cross-sections, number of generated events  and effect of cuts (C1 - C5) for the signal and 
relevant background processes . 
  Second
 column shows the cross-sections of respective processes in pb.
 Column, marked with $N_{EV}$, shows total the number of events
 generated for our analysis, subjected to the selection criteria
 defined in the text. Successive columns (marked with C1 - C5) show
 the remaining number of events after the application of the corresponding cut, 
 for signal and background processes. Here, P7, in the first row,
 corresponds to mUED parameters $R^{-1} = ~700$ GeV and $\Lambda R =
 10$. In the table $'*'$ indicates the background rate is negligible.  }
\label{table-cuts}       
	  \end{table}

Let us discuss the effects of the above cuts on the signal and
background.  More than 90\% (70 \%) of QCD1 (QCD2) jets $+ \etslash$
events are removed by C2. Remaining events are taken care by
application of C3 and C4. There is no real source of missing energy in
QCD processes. The missing energy in these events arise mainly from
the jet energy mis-measurements. As a result a cut of 50 GeV could
kill a substantial part of this background. C3 and C4 play the pivotal
role to reduce the $t \bar t$, W/Z $+$ jets events to a negligible
level.In addition the veto against tagged $b$-jets further reduce the $t
\bar t$ events. We have summarised the effects of the cuts in Table. 
\ref{table-cuts} .

We present the main results of our analysis in Table 2.  The number of 
events after all cuts for 1 fb$^{-1}$ luminosity, are presented in 
Table 2, for $R^{-1}$ values starting from 400 GeV upto 850 GeV in steps
 of 50 GeV (we 
denote these parameter points by P1, P2,......,P10) with two values of 
$\Lambda R = 10 ~{\rm and} ~40$.

\begin{table}[!htb]
\begin{center}\

\begin{tabular}{|c|c|c|c|c|c|c|c|c|c|c|}
\hline
\hline
          & P1    & P2   & P3   & P4   & P5  & P6  & P7  & P8   & P9   & P10  \\ 
\hline
$R^{-1}$      & 400   & 450  & 500  & 550  & 600 & 650 & 700 & 750  & 800  & 850  \\
\hline 
$\sigma_{10}$ & 116.2 & 55.4 & 28.6 & 15.3 & 8.4 & 4.8 & 2.8 & 1.64 & 1.01 & 0.59 \\
\hline
$\sigma_{40}$ & 83.5  & 40.3 & 20.3 & 10.7 & 5.8 & 3.2 & 1.9 & 1.08 & 0.64 & 0.38 \\
\hline
$(\sigma \times \epsilon)_{10}$ & 17.4&12.6&10.30 & 4.60&3.95 &3.02 & 2.52& 1.32 & 1.01 & 0.72 \\
\hline
$(\sigma \times \epsilon)_{40}$ & 23.4&17.3 &14.82 & 8.35&5.28 &3.71 & 2.68& 2.16 & 1.03 & 0.69 \\
\hline

\end{tabular}
\end{center}
\caption{Cross-sections for different representative parameter points
 in mUED model. Here $R^{-1}$ is in GeV. $\sigma_{10}$ and
 $\sigma_{40}$ denote the total cross-sections (in pb) from $G^1 G^1$,
 $G^1 Q^1$ and $Q^1 Q^1$ production for $\Lambda R =10$ and $\Lambda R
 =40$ respectively. $(\sigma \times \epsilon)_{10, 40}$ in 3rd and 4th
 row  denote jets $+ \etslash$ cross-sections (in fb) subjected to the
 cuts C1 - C5, from mUED model (for different values of R$^{-1}$) for 
 $\Lambda R =10$ and $\Lambda R
 =40$ respectively.}
\end{table}

As the SM background events have  been reduced to negligible 
levels, 10 signal events could be a potentially good number for the 
discovery. It is evident from the table that, with an accumulated 
luminosity of 12 fb$^{-1}$ (could be easily attainable by the end of 7 
TeV run of the LHC), mUED model can easily be probed via the jets $+ 
\etslash$ channel upto $R^{-1}$ of 850 GeV. However, even at 5 
fb$^{-1}$ integrated luminosity such signal can be probed upto $R^{-1} = 
$ 700 GeV.

At this point it is worthwhile to compare our results with two other
similar analyses \cite{kirtiman_biplob_ued, nojiri}, involving signals
containg one or more leptons, on exploring mUED at the LHC running at
7 TeV.  Analysis presented in \cite{kirtiman_biplob_ued}, has used the
conventional weapons of visible $p_T$ and $\etslash$ distributions to
fight with the SM backgrounds.  However, authors in
ref. \cite{kirtiman_biplob_ued}, used the multi-lepton (2- and
3-leptons) final states in association with jets (using 2 fb$^{-1}$
data at 7 TeV run of LHC) , to look for the mUED signal.  Assuming 5
events to be the benchmark for discovery for a background free signal,
the $R^{-1}$ reach in this paper, is in the ballpark of 700 GeV, with
2 fb$^{-1}$ of data. According to Ref. \cite{kirtiman_biplob_ued}, the
best reach is obtained in the tri-lepton ($+$ jets) channel.  This is
somehow expected, as the SM background rate in this channel is
practically vanishing.  Mass reach obtained in
Ref.\cite{kirtiman_biplob_ued} is also very similar to what has been
obtained in our analysis.  In another recent work \cite{nojiri},
authors have used a somewhat new strategy to explore the mUED
parameter space again at 7 TeV run of LHC. Here kinematic varible
$M_{T2}$ has been used to dig out the {\em 1 lepton $+$ jets} signal
arising from mUED, from the SM background.  However, projected mass
reach with 2 fb$^{-1}$ luminosity ($R^{-1}=$ 550 GeV with $\Lambda R$
= 10 and $R^{-1} = $ 600 GeV with $\Lambda R =$ 40) in our analysis is
certainly better than that ($R^{-1}$ = 400 GeV with $\Lambda R =$ 10
and $R^{-1} =$ 500 GeV with $\Lambda R =$ 40) presented in
Ref.\cite{nojiri}.

\section{Conclusion}

To summarise, we have explored the possibility of discovering the mUED
model at the LHC using the jets $+ \etslash$
channel, which among various signatures of mUED has the largest cross
section. It is well known that the mass splittings among different $n
=1$ KK-excitations of the SM particles are generically small as they
are generated by loop driven effects. As a result, typical
signatures of mUED would involve relatively low $p_T$ leptons and/or
jets accompanied by a soft $\etslash$ spectrum (see
Fig. \ref{dist_etmis}). In contrast, in nSUGRA motivated SUSY models the
corresponding signals consist of jets, leptons and $\etslash$ which
are considerably harder. Thus the traditional strong cuts on visible
or $\etslash$ which are often useful in isolating SUSY and other
new physics signals from the SM backgrounds, may not be very efficient
while searching for $n=1$ KK excitations in mUED.

For final states involving multiple leptons of moderately large $p_T$
signals of mUED may still be viable both at the LHC at 7 TeV
\cite{kirtiman_biplob_ued, nojiri} and 14 TeV \cite{BDMR,
kirtiman_ad_jhep} runs. However, the jets + $\etslash$ signal with the
largest cross-sections did not receive the due attention because of the
apprehension that in the absence of the coventional strong cuts, this
signal will be swamped by a large QCD background.
   
We, however, feel that this  signature having the largest
cross-sections, should be looked for at the LHC  for a complete 
understanding of the mUED model. To this end we have proposed   
a new search strategy. In view of our generator level simulations
it appears that even in the absence of the standard cuts usually 
employed for establishing new physics signals, a healthy signal in
the above channel can be established by a judicious use of the
event shape variables $\alpha_T$ and $R_T$.

We have generated the jets + $\etslash$ signal in mUED using
PYTHIA. The SM backgrounds have been estimated using ALPGEN and
PYTHIA. As expected attempts to remove the SM background by applying
strong cuts on $p_T$ of the jets and $\etslash$ , turned out to be
futile (see Figs. \ref{dist_etmis}). On the other hand demanding
$\alpha_T$ to be greater than 0.56 has eventually removed all the QCD
and W/Z + jets backgrounds. Additionally, demanding $R_T$ to be less
than 0.85 completely killed the $t \bar t$ and residual W/Z + jets
events (see Table 1). Requiring 10 signal events after all cuts is
then sufficient to claim a discovery for this background free
signal. We find that in mUED, $R^{-1}$ upto 850 GeV (700 GeV) can be
probed at the ongoing LHC experiments with 7 TeV center of mass energy
with an integrated luminosity of 12 fb$^{-1}$ (5 fb$^{-1}$) (Table 2).
Looking at the present performance of the LHC experiments, it may be
expected that this amount of data will be available by the end of 7
TeV run.

Though, we have demonstrated the utility of the event shape variables
in the context of mUED, these variables can as well be used for
searching a large class of new physics scenarios with compressed mass
spectra.

A case in point is the unconstrained minimal 
supersymmetric standard model (MSSM) with a mass 
difference of a few hundred GeV seperating  the
heaviest strongly interacting superparticle and the lightest 
supersymmetric particle (LSP). It can be readily checked that the $p_T$
distributions and the $\etslash$ distribution in a typical SUSY signal
in such a scenario will be relatively soft. Consequently the signal will 
be rather insensetive to the SUSY searches by the ATLAS and the CMS 
collaborations even if the squark-gluino masses are relatively small,
and cannot be constrained by the present LHC data.
It will be interesting to develop an alternative search strategy 
based on the event shape variables for these models. 

It may be recalled that it was pointed out long ago \cite{cms} that
the signatures of mUED and R-parity conserving mSUGRA could be
similar. However, in most versions of the MSSM like mSUGRA, the
sparticle spectra are well spread out and standard hard cuts can
separate the MSSM signal from the signatures of mUED. However, the
compressed version of the MSSM will indeed give rise to signals very
similar to the signals of mUED. It would then especially challenging
to differentiate between this compressed SUSY with mUED in the jets +
$\etslash$ channel . Event shape variables  may play a crucial role to
this end.

Several authors have discussed 
\cite{comprsd-SUSY, stealth-SUSY} the possibility of 
partially compressed spectra in the framework of supersymmetry
for various theoretical reasons.  
Characteristic signals at the LHC of such compressed spectra in  
mSUGRA type scenarios have also been discussed  
\cite{comprsd-SUSY}.
However, it should be noted that in neither of the 
models discussed  above the mass spectrum is as  
compressed as in the mUED model. Consequently, 
exploration/exclusion of such models at the LHC, can still be 
possible using  large visible/missing energy cuts.  However, it would 
be interesting to see whether the event shape variables can extend the 
mass reach at the LHC in these cases.

\vspace*{.5in}
\noindent
{\bf Acknowledgements :} Research of Anindya Datta is partially supported by the UGC-DRS programme at the Department of Physics, University of Calcutta.



\begin{thebibliography}{99}
\bibitem{ADD}I.~Antoniadis,
  Phys.\ Lett.\ B {\bf 246} (1990) 377;
N.~Arkani-Hamed, S.~Dimopoulous and G.~Dvali, {
  Phys.~Lett.~}{B~}{\bf 429}  (1998) 263; I.~Antoniadis,
  N.~Arkani-Hamed, S.~Dimopoulos and G.~R.~Dvali, {
  Phys.~Lett.~}{B~}{\bf 436}  (1998) 257.
  \bibitem{RS}L.~Randall and R.~Sundrum, { Phys.~Rev.~Lett.~}{\bf
83}  (1999) 3370; {\em ibid} {\bf 83}  (1999) 4690.
\bibitem{acd}T.~Appelquist, H.~C.~Cheng and B.~A.~Dobrescu,
  Phys.\ Rev.\ D {\bf 64} (2001) 035002
  [arXiv:hep-ph/0012100];  
  
  \bibitem{cms} H.~C.~Cheng, K.~T.~Matchev and M.~Schmaltz,
  Phys.\ Rev.\  D {\bf 66} (2002) 056006
  [arXiv:hep-ph/0205314].

\bibitem{antoniadis1} I.~Antoniadis, Phys.\ Lett.\ B {\bf 246} (1990) 377.
\bibitem{relax} G.~Bhattacharyya, S.~K.~Majee and A.~Raychaudhuri,
  Nucl.\ Phys.\  B {\bf 793} (2008) 114
  [arXiv:0705.3103 [hep-ph]].
\bibitem{Arkani-Hamed:1999dc} N.~Arkani-Hamed and M.~Schmaltz,
  Phys.\ Rev.\ D {\bf 61} (2000) 033005, [arXiv:hep-ph/9903417].
\bibitem{Servant:2002aq} G.~Servant and T.~M.~P.~Tait,
  Nucl.\ Phys.\ B {\bf 650} (2003) 391
  [arXiv:hep-ph/0206071].
\bibitem{Arkani-Hamed:2000hv} N.~Arkani-Hamed, H.~C.~Cheng, B.~A.~Dobrescu and L.~J.~Hall,
  Phys.\ Rev.\ D {\bf 62} (2000) 096006
  [arXiv:hep-ph/0006238].
\bibitem{dienes} K.~R.~Dienes, E.~Dudas and T.~Gherghetta,
  Phys.\ Lett.\ B {\bf 436} (1998) 55
  [arXiv:hep-ph/9803466]; K. Dienes, E. Dudas, and T. Gherghetta; Nucl. \ Phys.\ B
{\bf 537} (1999) 47   [arXiv:hep-ph/9806292]; For a parallel analysis based on a minimal length scenario, see
  S.~Hossenfelder,
  Phys.\ Rev.\ D {\bf 70} (2004) 105003
  [arXiv:hep-ph/0405127].
\bibitem{blitz}G.~Bhattacharyya, Anindya ~Datta, S.~K.~Majee and A.~Raychaudhuri,
  Nucl.\ Phys.\  B {\bf 760} (2007) 117
  [arXiv:hep-ph/0608208].
\bibitem{rad_corr}
M.~Puchwein and Z.~Kunszt,
Annals Phys.\ {\bf 311} (2004) 288 [arXiv:hep-th/0309069];
H.~Georgi, A.~K.~Grant and G.~Hailu,
Phys.\ Lett.\ B {\bf 506} (2001) 207
[arXiv:hep-ph/0012379];
G.~von Gersdorff, N.~Irges and M.~Quiros,
Nucl.\ Phys.\ B {\bf 635} (2002) 127
[arXiv:hep-th/0204223].


\bibitem{BDMR}G.~Bhattacharyya, Anindya ~Datta, S.~K.~Majee and A.~Raychaudhuri,
  Nucl.\ Phys.\  B {\bf 821} (2009) 48, 
  [arXiv:hep-ph/0608208].

\bibitem{db}P.~Dey and G.~Bhattacharyya,
  Phys.\ Rev.\ D {\bf 70} (2004) 116012
  [arXiv:hep-ph/0407314];
P.~Dey and G.~Bhattacharyya,
  Phys.\ Rev.\ D {\bf 69} (2004) 076009
  [arXiv:hep-ph/0309110].

\bibitem{nath}P.~Nath and M.~Yamaguchi,
Phys.\ Rev.\ D {\bf 60} (1999) 116006
[arXiv:hep-ph/9903298].
\bibitem{chk}D.~Chakraverty, K.~Huitu and A.~Kundu,
Phys.\ Lett.\ B {\bf 558} (2003) 173
[arXiv:hep-ph/0212047]; A.J.~Buras, M.~Spranger and A.~Weiler,
Nucl.\ Phys.\ B {\bf 660} (2003) 225
[arXiv:hep-ph/0212143];
A.J.~Buras, A.~Poschenrieder, M.~Spranger and A.~Weiler,
Nucl.\ Phys.\ B {\bf 678} (2004) 455
[arXiv:hep-ph/0306158];
K.~Agashe, N.G.~Deshpande and G.H.~Wu,
Phys.\ Lett.\ B {\bf 514} (2001) 309
[arXiv:hep-ph/0105084].

\bibitem{santa}
  J.~F.~Oliver, J.~Papavassiliou and A.~Santamaria,
  Phys.\ Rev.\  D {\bf 67} (2003) 056002
  [arXiv:hep-ph/0212391].

\bibitem{appel-yee}
  T.~Appelquist and H.~U.~Yee,
  Phys.\ Rev.\ D {\bf 67} (2003) 055002
  [arXiv:hep-ph/0211023].


\bibitem{ewued} T.G. Rizzo and J.D. Wells, Phys. Rev. D {\bf 61}
(2000) 016007 [arXiv:hep-ph/9906234]; A. Strumia, Phys. Lett. B {\bf
466} (1999) 107 [arXiv:hep-ph/9906266]; C.D. Carone, Phys. Rev. D {\bf
61} (2000) 015008 [arXiv:hep-ph/9907362].

\bibitem{collued} T. Rizzo, Phys. Rev. D {\bf 64} (2001) 095010
[arXiv:hep-ph/0106336]; C. Macesanu, C.D. McMullen and S. Nandi,
Phys. Rev. D {\bf 66} (2002) 015009 [arXiv:hep-ph/0201300];
Phys. Lett. B {\bf 546} (2002) 253 [arXiv:hep-ph/0207269];
H.-C. Cheng, Int. J. Mod. Phys. A {\bf 18} (2003) 2779
[arXiv:hep-ph/0206035]; A. Muck, A. Pilaftsis and R. R\"uckl,
Nucl. Phys. B {\bf 687} (2004) 55 [arXiv:hep-ph/0312186];
  B.~Bhattacherjee and A.~Kundu,
  J.\ Phys.\ G {\bf 32}  (2006) 2123
  [arXiv:hep-ph/0605118];
  B.~Bhattacherjee and A.~Kundu,
  Phys.\ Lett.\  B {\bf 653} (2007) 300
  [arXiv:0704.3340 [hep-ph]];
  P.~Bandyopadhyay, B.~Bhattacherjee and Aseshkrishna ~Datta,
  JHEP {\bf 1003}  (2010) 048
  [arXiv:0909.3108 [hep-ph]];
  B.~Bhattacherjee, A.~Kundu, S.~K.~Rai and S.~Raychaudhuri,
  Phys.\ Rev.\ D {\bf 81}  (2010) 035021
  [arXiv:0910.4082 [hep-ph]].
  \bibitem{kong1}
  Aseshkrishna ~Datta, K.~Kong and K.~T.~Matchev,
  Phys.\ Rev.\  D {\bf 72} (2005) 096006.
  [hep-ph/0509246].

\bibitem{kirtiman_ad_jhep}D.~Choudhury, Anindya ~Datta and K.~Ghosh,
  JHEP {\bf 1008} (2010) 051
  [arXiv:0911.4064 [hep-ph]].
\bibitem{kirtiman_biplob_ued}B.~Bhattacherjee, K.~Ghosh,
  Phys.\ Rev.\  D{\bf 83} (2011) 034003.
  [arXiv:1006.3043 [hep-ph]].
\bibitem{ILC}G.~Bhattacharyya, P.~Dey, A.~Kundu and A.~Raychaudhuri,
  Phys.\ Lett.\  B {\bf 628} (2005) 141
  [arXiv:hep-ph/0502031];
 Anindya ~Datta and S.~K.~Rai,
  Int.\ J.\ Mod.\ Phys.\  A {\bf 23} (2008) 519 
  [arXiv:hep-ph/0509277];
  B.~Bhattacherjee, A.~Kundu, S.~K.~Rai and S.~Raychaudhuri,
  Phys.\ Rev.\  D {\bf 78} (2008) 115005 
  [arXiv:0805.3619 [hep-ph]];
  B.~Bhattacherjee,
  Phys.\ Rev.\  D {\bf 79} (2009) 016006 
  [arXiv:0810.4441 [hep-ph]].

\bibitem{kong}M.~Battaglia, Aseshkrishna ~Datta, A.~De Roeck, K.~Kong and K.~T.~Matchev,
  JHEP {\bf 0507} (2005) 033.
  [arXiv:hep-ph/0502041]; 
B.~Bhattacherjee and A.~Kundu,
  Phys.\ Lett.\  B {\bf 627} (2005) 137
  [arXiv:hep-ph/0508170].
  
\bibitem{SUSY-CMS} S. Chatrchyan et al., CMS Collab., arXiv:1109.2352 ; CMS-SUS-11-003 ; CERN-PH-EP-2011-138. - 2011.
\bibitem{SUSY-ATLAS}ATLAS Collab., CERN-PH-EP-2011-145. - 2011. 
\bibitem{comprsd-SUSY}T. J. LeCompte,  S. P. Martin,  Phys. Rev. {D84} (2011) 015005.
\bibitem{stealth-SUSY}J. J. Fan, M. Reece and J. T. Ruderman, arXiv:1105.5135 [hep-ph].\bibitem{nojiri}H. ~Murayama, M. ~Nojiri and K. ~Tobioka,  [arXiv:1107.3369 [hep-ph]].
\bibitem{LKP_dark_matter} K.~Kong, K.~T.~Matchev,
  JHEP {\bf 0601} (2006) 038.
\bibitem{PYTHIA}
T. Sjostrand, P. Eden, C. Friberg, L. Lonnblad, G. Miu,
S. Mrenna and  E. Norrbin, Comp. Phys. Comm. {\bf 135} (2001) 238, For a more recent version see,
[arXiv: hep-ph/0603175].
\bibitem{CTEQ6} J. Pumplin {\em et al.}, JHEP {\bf07} (2002) 012.
\bibitem{ALPGEN} M. Mangano {\em et al.}, JHEP {\bf 0307} (2003) 001.
\bibitem{Guchait} M. Guchait, D. Sengupta,  arXiv:1102:4785 [hep-ph].
\bibitem{randall-alpha-T} L. Randall, D. Tucker-Smith, Phys.Rev.Lett. {\bf 101} (2008) 221803.
\bibitem{cms_trigger}V. Khachatryan et al., CMS Collab., Phys.\ Lett.\  B {\bf 698} (2011) 196;  arXiv:1101.1628[hep-ex] . 

\end{thebibliography}
\end{document}